\documentclass[12pt]{article}

\usepackage{paper2e}
\usepackage{mydefs2e}
\usepackage{xspace}
\usepackage{graphicx}

\newcommand{\gap}{\hspace{0.05em}}
\newcommand{\met}{\,\sla{\!E}_T}

\begin{document}

\begin{titlepage}

\title{Asymmetric Dark Matter}

\author{David E. Kaplan}

\address{Physics Department, Johns Hopkins University\\
Baltimore, Maryland 21218}

\author{Markus A. Luty}

\address{Physics Department, University of California\\
Davis, California 95616}

\author{Kathryn M. Zurek}

\address{Center for Particle Astrophysics,
Fermi National Accelerator Laboratory \\
Batavia, IL  60510 \\
$\vphantom{\biggl[}$ and \\
Physics Department, University of Wisconsin\\
Madison, Wisconsin 53706 \\
}

\begin{abstract}
We consider a simple class of models in which the relic density of
dark matter is determined by the baryon asymmetry of the universe.
In these models a $B - L$ asymmetry generated at high temperatures
is transfered to the dark matter, which is charged under $B - L$.
The interactions that transfer the asymmetry decouple at temperatures
above the dark matter mass, freezing in a dark matter asymmetry
of order the baryon asymmetry.
This explains the observed relation between
the baryon and dark matter densities for dark matter mass in the
range $5$--$15$ GeV.
The symmetric component of the dark matter can annihilate
efficiently to light pseudoscalar Higgs particles $a$,
or via $t$-channel exchange of new scalar doublets.
The first possibility allows for $h^0 \to aa$ decays,
while the second predicts a light charged Higgs-like
scalar decaying to $\tau\nu$.
Direct detection can arise from Higgs exchange in the first model,
or a nonzero magnetic moment in the second.
In supersymmetric models, the would-be LSP can decay into pairs
of dark matter particles plus standard model particles,
possibly with displaced vertices.
\end{abstract}

\end{titlepage}

\section{Introduction}
\label{sec:intro}

There is compelling evidence from astrophysical and cosmological
data that the dominant contribution of the matter in the universe
is in the form of ``dark matter'' that interacts very
weakly with ordinary matter \cite{SilkReview}.
One of the striking features of this picture is that the
dark matter density today is rather close to the baryon density:
$\rho_{\rm DM} \simeq 4.5 \rho_{\rm baryon}$
in the standard cosmological model \cite{cosmomodelreview},
suggesting that these relic densities
have a common origin.
However, in the standard paradigm for dark matter,
the dark matter and baryon relic densities arise by completely
different mechanisms, and the fact that they have the same
order of magnitude is a ``cosmic coincidence.''

In the standard cosmology the baryon relic density arises
from a tiny baryon-antibaryon asymmetry of order $10^{-10}$ 
at temperatures above $10\MeV$.
This paradigm is strongly supported by the success of
big-bang nucleosythesis.
The baryon asymmetry can be generated starting from
an initially symmetric universe (``baryogenesis'')
if baryon number and $C\!\gap P$ are
violated out of equilibrium in the early universe \cite{Sakharov}.
Non-perturbative effects in the Standard Model efficiently
violate baryon and lepton number at temperatures above
the electroweak phase transition ($T \gsim 100\GeV$),
so the simplest possibility is that a $B - L$ asymmetry
is generated at high scales,
\eg\ by leptogenesis.

In contrast with the baryon relic density, the origin of the
dark matter abundance is not strongly constrained by cosmological data.
The most popular model is a weakly-interacting massive particle
(WIMP) whose relic density is determined by the freeze-out of
its annihilations to standard model particles.
This naturally explains the observed order of magnitude of the dark
matter relic abundance, but not why this is close to the
baryon abundance.

In this paper we consider a simple explanation for
$\rho_{\rm DM} \sim \rho_{\rm baryon}$,
namely that the dark matter density arises from
a dark matter particle-antiparticle asymmetry
related to the $B - L$ asymmetry.
Previous models based on this idea are described in
\cite{Kaplan,Barr,Dodelson,Fujii,Kitano,Farrar}.
In our models, the $B - L$ and dark matter asymmetries
can be related by interactions in
equilibrium that transfer the $B - L$ asymmetry 
(assumed to arise from a standard baryogenesis mechanism)
to the dark matter.
Any interaction that forces the dark matter to carry a nonzero
$B - L$ charge will accomplish this.
Since the dark matter relic density is set by the baryon asymmetry
and not by the properties of thermal freeze-out,
we term this class of models Asymmetric Dark Matter (ADM).

This mechanism predicts $n_{\rm DM} \sim n_{B}$,
and therefore
$\Om_{\rm DM} \sim (m_{\rm DM}/m_B) \Om_B$.
We therefore obtain the observed dark matter abundance for
$m_{\rm DM} \sim 5\GeV$.
The precise dark matter mass is calculable in a given
model, and the models we construct give values in the range
from $5$ to $15\GeV$.
These values are close to the electroweak scale, suggesting
that the dark matter mass is generated
by electroweak symmetry breaking.
This also gives a possible mechanism for the annihilation of the
symmetric component of the dark matter as well as a direct detection
mechanism.

Dark matter masses in this range may give
an explanation of the 
DAMA observations \cite{DAMA,GG,Fornengo,Petriello}.
The DAMA experiment has observed an annual modulation with
$8.2\sigma$ significance
consistent with WIMP scattering.
Other direct dectection experiments are sensitive to lower
cross sections, but have higher energy thresholds, so a
WIMP in the 10\GeV\ mass range may explain the DAMA signal while
still being consistent with other null results \cite{Petriello,Savage}.
There remains some controversy, however, since the region consistent with
the DAMA signal depends on the choice of binning \cite{otherfits}.
The fit is particularly sensitive to the 2--$2.5\keV$ nuclear recoil bin;
this bin taken by itself tends to shift the fit to larger dark matter
masses \cite{sundama}.
We therefore emphasize that the models discussed here are
interesting independently of the motivation from the DAMA
observation.

The main features of our models are as follows.
\begin{itemize}
\item 
At a high temperature, a $B-L$ asymmetry is generated.
Below this temperature $B-L$ is preserved,
but new higher-dimension effective
interactions that exchange $B-L$ number between the
standard model and dark matter are in thermal equilibrium.

\item The interactions that exchange the $B-L$ asymmetry decouple
at lower temperatures,
and a dark matter asymmetry is frozen in.
There may still be couplings in equilibrium between the
dark matter and the standard model that do not transfer the
asymmetry.

\item 
At temperatures below the mass of the dark matter particle,
particle-antiparticle annihilations eliminate the
symmetric component of the dark matter density,
leaving behind a relic density proportional to the particle-antiparticle
asymmetry.
These annihilations may occur either through the interactions
that generate the dark matter mass
or via the operator that transfers the asymmetry.

\item
Direct detection of dark matter
may occur, also through the interactions that generate the
dark matter mass
or via the operator that transfers the asymmetry.

\item
In supersymmetric models, the dark matter particle is
naturally the lightest particle charged under a discrete
$R$ symmetry, and the would-be LSP decays to pairs of
dark matter particles plus standard model particles.
These decays may have a macrosopically displaced vertex.
\end{itemize}

This paper is organized as follows.
In Section 2, we describe concrete models
and explain in detail how they give rise to the observed
dark matter density.
In Section 3, we discuss direct detection signals.
In Section 4, we discuss novel collider signals
in this class of models.
Section 5 contains our conclusions.

\section{Models}
It is simple to construct specific models that generate ADM,
and we will give three examples below.
We find it simplest to explain the details of the mechanism in terms of
a specific ``reference'' model.
The remaining models will be described more briefly.

\subsection{Reference Model: $L = \frac 12$ ADM}
\label{RefModel}
We begin with a supersymmetric model in which the dark
matter carries lepton number.
Supersymmetry (SUSY) is not necessary for
the dark matter mechanism we are studying,
but it allows a direct connection to a realistic and compelling
model of electroweak physics, and leads to very interesting
collider phenomenology.
Before going into the details of the model,
we outline its general features:
\begin{itemize}
\item 
The dark matter sector consists of a pair of
gauge singlet chiral superfields $X$, $\bar{X}$
with $L = \pm \frac 12$.
This allows a supersymmetric mass term of the form $\bar{X}X$.
There may be $\De L = 2$ breaking of lepton number from Majorana
neutrino masses, but a
$Z_4$ subgroup of $U(1)_L$ remains unbroken.
This forbids Majorana mass terms of the form $X^2$ and
$\bar{X}^2$ that can efficiently wipe out the asymmetry,
and also guarantees that the lightest component of 
$X$ is a stable dark matter candidate.

\item
A $B - L$ asymmetry generated at high scales
is transfered to the dark matter via
the effective interaction
\beq[LeffSUSY1]
\De W_{\rm eff} =
\frac{1}{M_i}
\bar{X}^2 L_i H_u, 
\eeq
where $M_i$ is a high mass scale parameterizing the new physics
that generates the interaction.
The lowest-dimension interactions allowed by the unbroken $Z_4$ subgroup
of $U(1)_L$ are dimension-5 operators of the form
$\De W \sim X^4$.
As long as these drop out of equilibrium at a temperature
where \Eq{LeffSUSY1} is still in equilibrium, the
asymmetry will be transfered to the visible sector.
The interaction \Eq{LeffSUSY1} naturally goes out of equilibrium
as the tempurature drops further,
and the dark matter asymmetry freezes in.

\item
The dark matter mass is close to the electroweak scale,
suggesting that it arises from electroweak symmetry breaking.
This can occur in the next-to-minimal supersymmetric model
(NMSSM), where the $X$ mass arises from the coupling to
a singlet $S$ that gets a VEV at the weak scale.

\item
The annihilation of the symmetric component of the dark matter thermal
density can occur through Higgs exchange in
the NMSSM, or via the interaction that transfers the asymmetry.

\item
Direct detection can occur through the same interactions that
are responsible for the annihilation.
Either can give a signal in upcoming experiments.

\item
The operator \Eq{LeffSUSY1} violates the usual $R$ parity,
but preserves a $Z_4$ $R$ symmetry under which the MSSM
fields have the usual $R$ parity assignments $\pm 1$ while
\beq
X(\th) \mapsto i X(-\th),
\qquad
\bar{X}(\th) \mapsto -i X(-\th),
\eeq
where $\th$ is the superspace coordinate.
The would-be LSP can therefore decay into pairs of $X$
particles via the operator \Eq{LeffSUSY1}.
The $Z_4$ $R$ symmetry forbids $R$-parity violating operators
from being generated in the visible sector.
\end{itemize}
\noindent
We now describe the model and the mechanism in more detail.

We begin by briefly discussing the UV completion of this model.
The effective interaction \Eq{LeffSUSY1} can be obtained from a theory
with a heavy pair of chiral multiplets $N$, $\bar{N}$ with
$L = \pm 1$, \ie\ vectorlike sterile neutrinos:
\beq[UVcomplete1]
\De W = M \bar{N} N +
\la' N \bar{X}^2 + y'_i \bar{N} L_i H_u.
\eeq
Another possibility is a vector-like pair of
electroweak doublets $D$ and $\bar{D}$ with
$L = \pm \frac 12$:
\beq[UVcomplete2]
\De W = M \bar{D} D + \la' \bar{X} D H_u
+ y'_i L_i \bar{D} \bar{X}.
\eeq
Either model generates \Eq{LeffSUSY1} with $M_i = M / (\la' y'_i)$.
Note that in the second model \Eq{UVcomplete2} the Higgs VEV gives a mixing
between $X$ and the neutral component of $D$;
this can be treated as a small perturbation as long as $M \gg \la' v_u$.
The second model is more natural if Majorana neutrino masses are
generated from a standard see-saw mechanism.
The reason is that the sterile neutrino Majorana mass 
in the standard see-saw
naturally arises from the VEV of a field with $L = 2$.
Such a field has no renormalizable couplings to the the fields
in \Eq{UVcomplete2}, and therefore lepton number can be
an accidental symmetry in the dark matter sector.
This also allows the $B - L$ asymmetry to be generated
by standard leptogenesis.

We now discuss the generation of the dark matter asymmetry
in this model.
In the early universe, non-renormalizable effective interactions
such as \Eq{LeffSUSY1} give rise to interactions whose rate
drops faster than the expansion rate as the temperature of the
universe drops.
It is therefore natural for this operator to have been in equilbrium
in the early universe, but be out of equilbrium today.
This is exactly what is required to distribute a $B - L$
asymmetry in the early universe to the $X$ fields.
This goes out of equilibrium for $T \gsim 100\GeV$ 
provided that $M \gsim 10^9\GeV$.
For this coupling strength, the bounds on induced lepton
flavor violation such as $\mu \to e \ga$ are well below
the experimental limits.

It is also possible for the interactions that transfer the asymmetry to
go out of equilibrium at temperatures below the electroweak scale but
above the dark matter mass even if $M \ll 10^9 \mbox{ GeV}$.
This is a small temperature range (roughly $10\GeV$
to $100\GeV$) but we will see this
arises naturally for a wide range of parameters.
The component interactions arising from \Eq{LeffSUSY1} contain at most
two fermion fields, so the only interactions that change the $X$
fermion number arising directly from \Eq{LeffSUSY1} involve
sneutrino and/or Higgs particles, \eg
$\tilde{\nu} \leftrightarrow \bar{X} \bar{X}$.
These interactions become ineffective at temperatures
below the sneutrino mass because of the exponentially small
abundance of sneutrinos.
The rate is
\beq[lowdecouplereal]
\Ga(\tilde\nu \leftrightarrow XX) \sim
\frac{n_{\tilde\nu}}{n_X}
\frac{1}{16\pi} \left( \frac{v_u}{M} \right)^2 m_{\tilde\nu}.
\eeq
This freezes out when the rate drops below the Hubble expansion
rate, which occurs for $T \lsim m_{\tilde\nu}/40$
for $M \sim \mbox{TeV}$.
In addition, there are transitions between light particles
generated by integrating out virtual heavy particles.
Since all light particles are neutral under the $Z_4$ $R$ symmetry,
we need two insertions of the operator \Eq{LeffSUSY1}.
The leading contribution arises from integrating out virtual
sneutrinos and neutralinos,
and gives rise to an effective operator
\beq[Leffeffmodel1]
\scr{L}_{\rm eff} \sim \frac{v_u^2}{M^2 m_{\tilde{\nu}}^4 m_{\tilde{B}}}
(\bar{X} \bar{X})^2 \nu \nu.
\eeq
The asymmetry-transferring processes mediated by this interaction 
(\eg\ $\bar{X}\bar{X} \leftrightarrow XX\bar\nu\bar\nu$)
have a rate that
falls rapidly as the universe cools:
\beq
\Gamma \sim \frac{1}{16 \pi} \left( \frac{1}{8 \pi^2} \right)^2
\left(\frac{v_u^2}{M^2 m_{\tilde{\nu}}^4 m_{\tilde{B}}}
\right)^2 T^{11},
\eeq
where the prefactor is an estimate of 4-body phase space.
This goes out of equilibrium for
\beq[lowdecouplevirtual]
T \lsim 20\GeV \left( \frac{M}{\mbox{TeV}} \right)^{4/9}
\left( \frac{m}{100 \mbox{ GeV}} \right)^{10/9},
\eeq
where $m \sim m_{\tilde\nu} \sim m_{\tilde B}$.
We see that for $m_X \sim 10\GeV$ the temperature where the interactions
decouple can be above $m_X$ even if $M$ is near the weak scale.
In this case, the interactions \Eq{Leffeffmodel1}
fall out of equilibrium before the dark matter becomes
non-relativistic, and the dark matter asymmetry is not washed out.%
\footnote{%
Sphaleron transitions may fall out of equilibrium in this temperature
range, but this does not have a large effect on the dark matter
asymmetry.
Below the sphaleron decoupling temperature $B$ and $L$ are
effectively separately conserved, but this does not prevent the
operator \Eq{LeffSUSY1} from transferring the asymmetry.}

For low values of $M$,
bounds from lepton flavor violation such as $\mu \to e\ga$ can be
satisfied if the coupling \Eq{LeffSUSY1}
is dominantly to third generation leptons.
This is what we expect if flavor symmetry is most badly broken
for the heavier generations.
For these low-scale models, the interaction \Eq{LeffSUSY1}
may provide both an annihilation and a direct detection mechanism,
which will be discussed in Section~3 below.

We now discuss the calculation of the $X$ 
particle-antiparticle asymmetry.
Because the asymmetry is transferred by interactions in
equilibrium, we can compute the $X$ asymmetry
in terms of the particle-antiparticle asymmetries of the standard
model using standard equilibrium methods \cite{HarveyTurner}.
The value of the $X$ asymmetry at low temperatures depends
on the temperature where the interactions \Eq{LeffSUSY1}
drop out of equilibrium.
We first discuss the case where these interactions
drop out of equilibrium above the electroweak
phase transition.
We then have
\beq
X = -\frac{11}{79} (B - L),
\eeq
where $X$ is the ``$X$ number'' charge.
$B - L$ is preserved by the electroweak phase transition,
so the present baryon asymmetry is proportional
to $B - L$.%
\footnote{%
We are assuming that there is no significant baryon asymmetry
generated during the electroweak phase transition.}
Sphaleron transitions that violate $B$ and $L$
are in equilibrium below the electroweak phase transition
\cite{sphalrate}.
The precise relation between $B$ and $B - L$ depends on
finite mass effects, \eg\ for the top quark.
Numerically, however, these do not make a large difference,
and we find
\beq
\frac{B}{B - L} \simeq 0.31.
\eeq
Assuming that the $X$ asymmetry is responsible for the observed
dark matter density then gives a prediction for the
mass of the $X$ particle if the interactions fall out of equilibrium
above the electroweak phase transition:
\beq[Xmassrefmodel]
m_X \simeq 2.4\GeV \,\,
\frac{\Om_{\rm DM}}{\Om_B} \simeq 11\GeV.
\eeq
The fact that the $X$ mass is somewhat larger than the
\naive\ estimate of $5\GeV$ is due to $X < B$,
which in turn can be traced to the fact that the model
contains more baryons than $X$ particles: in
relativistic equilibrium conserved
charges are proportional to the number of degrees of
freedom carrying that charge.%
\footnote{We must also impose the condition that the
universe has no net electric charge.
Since $X$ does not carry charge, this condition restricts
only the relative number of standard model particles, and does
not affect the scaling argument above.}

It is also possible that the interactions \Eq{LeffSUSY1}
decouple below the electroweak phase transition.
In this case, integrating out both the top and the superpartners, we obtain
\beq
\frac{X}{B} = \frac{13}{40}
\eeq
and therefore
\beq
m_X \simeq 13\GeV.
\eeq	

We now discuss the origin of the dark matter mass.
This is a supersymmetric Dirac mass arising from a
superpotential term $\De W = m_X \bar{X} X$.
The question of why $m_X$ is close to the weak scale is
similar to the ``$\mu$ problem'' of supersymmetric models,
which is explaining the origin of the supersymmetric Higgs
mass term
$\De W_{\rm eff} = \mu H_u H_d$.
Perhaps the simplest solution is 
the next-to-minimal supersymmetric standard model (NMSSM)
in which the required mass terms are given
by the VEV of a singlet field $S$:
\beq[NMSSM]
\De W = 
\la_X S X \bar{X}
+ \la_H S H_u H_d + \frac{\ka}{3} S^3. 
\eeq
This model naturally generates a VEV for $S$ of order the
electroweak scale and gives the required mass terms for
Higgs and $X$ particles.
Very importantly for dark matter phenomenology, it also
gives a direct coupling of $X$ to the standard model,
allowing the dark matter to be directly detected.

The final ingredient is that the thermal abundance of $X$ particles and antiparticles must efficiently
annihilate, so that the relic density of dark matter is
given by the $X$ particle-antiparticle asymmetry.
This requires
$\avg{\si_{\rm ann} v} \gsim \mbox{pb}$.
In the context of the NMSSM, a simple possibility is
$\bar{X}X \to aa$, where $a$ is the lightest pseudoscalar
in the Higgs sector.
This is unsuppressed in the early universe as long as
$m_a \lsim m_X$.
It is natural for $a$ to be light if $A$
terms are small, in which case $a$ is a pseudo Nambu-Goldstone
boson of a global $U(1)_R$ symmetry.
The annihiation comes from the coupling
\beq
\De\scr{L}_{\rm eff} = m_X \bar{X} X e^{ia/s} + \hc,
\eeq
where $s/\sqrt{2} = \avg{S}$,
which gives an annihilation cross section
\beq
\avg{\si v_{\rm rel}} = \frac{1}{16\pi} \frac{m_X^2}{s^4}.
\eeq
This is larger than $1~\mbox{pb}$ for $s < 200\GeV$.
This requires superpotential couplings
$\la, \ka \sim \scr{O}(1)$
to generate the correct spectrum in the NMSSM, and the resulting theory
is not perturbative up to the GUT scale.
This means that an extension of the Higgs sector is required
at high scales, such as in ``fat'' Higgs models \cite{fatHiggs}.
The pseudoscalar $a$ can decay promptly to $\bar{b}b$
(for $m_a \gsim 10\GeV$)
or $\tau^+ \tau^-$ (for $2\GeV \lsim m_a \lsim 10\GeV$),
so there are no further cosmological consequences.
Interestingly, this model points to the same region of NMSSM
parameter space where non-standard Higgs decays
such as $h^0 \to aa$ followed by
$a \to \bar{b}b$ or $\tau^+\tau^-$ can dominate,
which may alleviate the naturalness problems of supersymmetry
\cite{Dermisek:2005ar}.
We will also see that this model may give rise to a direct detection
signal.

Another possibility for annihilation is that the singlet couples
weakly to the Higgs fields.
In this case the $\mu$ term is not explained by the VEV of
$S$; it may arise \eg\ by the Giudice-Masiero mechanism \cite{GM}.
The theory can be perturbative up to the GUT scale
without additional structure.
The light $a$ can result from an approximate $U(1)_R$ symmetry
acting only on $S$, and the decay $a \to \bar{b}b$ or $\tau^+\tau^-$
need only be faster than a second to avoid constraints from
nucleosynthesis.
This annihilation mechanism does not give any observable direct
detection or collider signals. 

Another possibility for annihilation arises from the fields 
in the UV completion \Eq{UVcomplete2} if the doublets $D$ and
$\bar{D}$ are light.
Assuming dominance of heavy flavors,
we then have the annihilation channels
$\bar{X}X \to \bar{\nu}\nu, \tau^+ \tau^-$
from $t$-channel exchange of the scalar component of the doublet 
with rate
\beq
\avg{ \si v_{\rm rel}}
= \frac{1}{16\pi} \frac{y'^4 m_X^2}{m_{\tilde{D}}^4} .
\eeq
This is larger than a pb for 
$m_{\tilde{D}} / y' < 190\GeV$.
The coupling $y'$ breaks lepton flavor symmetry, and
suppressing lepton flavor-violating processes such as $\mu \to e\ga$
requires nontrivial structure in the lepton flavor sector.
For example, there may be an approximate $U(1)_L^3$ forbidding
lepton mixing that is broken only by small neutrino masses.
The charged doublet scalar can be pair produced and decays to
$\tau^\pm + \,\sla{\!E}_T$.
LEP bounds give $m_{\tilde{D}^\pm} > 92\GeV$ \cite{LEPcharged};
there is currently no Tevatron search for this mode.
In order for the operator \Eq{LeffSUSY1} that transfers the asymmetry
to decouple at temperatures
above $m_X$ we need $\la' \ll 1$ in \Eq{UVcomplete2}.
Note that this also suppresses the mixing between $X$ and the neutral
component of the doublet, which would otherwise lead to coupling of
$X$ to the $Z$.
Having $\la ' \ll 1$ is natural because in the limit $\la' \to 0$
there is an enhanced global $U(1)$ symmetry under which $X$ and $D$
have opposite charges.
As we will discuss in Section 3, this model has a direct detection
cross section via a charge radius interaction that is near the
experimental limit, making this version of the mode phenomenologically
very interesting.

There are other possibilities for the annihilation,
such as annihilation into light hidden sector fields
or other couplings to standard model fields, \eg\ via a $Z'$.
Another interesting possibility to explore are models where
annihilation occurs via new strong dynamics, as in
``hidden valley'' \cite{HV} or ``quirk'' \cite{quirk} models.
We leave these possibilities for future work.

\subsection{$B = \frac 12$ ADM}
\label{Model2}
We now describe a supersymmetric model in which the dark
matter carries baryon number.
The model is a very simple variation of the previous
model, so our discussion will be very brief.
The model consists of the MSSM plus a pair of gauge singlet
chiral superfields $X$, $\bar{X}$ with $B = \pm \frac 12$.
The lowest-dimension operator that can transfer the baryon
asymmetry to $X$ is
\beq[LeffSUSY2]
\De W_{\rm eff} = 
\frac{1}{M_{ijk}^2} \bar{X}^2 u_i d_j d_k.
\eeq
If this interaction goes out of equilibrium above the
electroweak phase transition, we find
\beq
X = -\frac{11}{79} (B - L).
\eeq
Amusingly, this is precisely the same result as in the previous
model, and we again find
\beq
m_X \simeq 11\GeV
\eeq
if the interaction \Eq{LeffSUSY2}
decouples above the electroweak phase transition.
The $\bar{X}X$ annihilation
and the generation of the $X$ 
mass are very similar to the previous model,
and we will not repeat the discussion.
A significant difference between this model and the $L=\frac 12$
model is the long lifetime of the LSP due to the high dimension of
the transfer operator.
As we will see in Section \ref{collide} below, the
scale $M$ in some cases must be of order a TeV or smaller in order
to avoid decays on cosmological time scales.

\subsection{$L = 1$ (Sterile Neutrino) ADM}
\label{NuADM}
We now consider a model in which the dark matter has
$L = 1$, like a sterile neutrino.
The lowest-dimension coupling to the standard model that
can transfer the lepton asymmetry to $X$ is then
\beq[LeffnonSUSY]
\De\scr{L}_{\rm eff} = \frac{1}{M_{ij}^4} \bar{X}^2 (L_i H) (L_j H)
+ \hc
\eeq
Majorana neturino masses conventionally arise from an effective
operator of the form $(LH)^2$, so
any model that generates the interaction \Eq{LeffnonSUSY}
and Majorana masses necessarily generates a Majorana
mass term for $X$ at some level.
This will efficiently wipe out any $X$ asymmetry, so this
model is most natural with Dirac neutrino masses.

If the interaction \Eq{LeffnonSUSY} goes out of equilibrium above the weak scale, we have
\beq
X = -\frac{12}{49} (B - L),
\eeq
corresponding to a dark matter particle mass $m_X \simeq 6\GeV$.

A UV completion of \Eq{LeffnonSUSY} can be obtained by
adding an additional scalar Higgs doublet $H'$ with couplings
\beq[LeffnonSUSYUVcomplete]
\De\scr{L} = y'_i L_i H' \bar{X} 
- \frac{\la'}{4} \left[ (H^\dagger H')^2 + \hc \right]
+ \cdots
\eeq
Integrating out $H'$ generates \Eq{LeffnonSUSY} with
$1/M_{ij}^4 \sim \la' y'_i y'_j / m_{H'}^4$.
Note that $H'$ is odd under the $Z_2$ symmetry that prevents
$X$ decay, so we must assume that $\avg{H'} = 0$.

Exchange of $H'$ can give rise to annihilation for the
symmetric component of the $X$ relic density if
$m_{H'}/y' \lsim 200\GeV$.
This is very similar to exchange of the doublet scalars
in the $L = \frac 12$ model.
It requires $\la' \ll 1$ in order to decouple the
transfer of the asymmetry above $m_X$,
and nontrivial lepton flavor structure to avoid
lepton flavor violation such as $\mu \to e\ga$.
Direct detection signals from the light $H'$ will be
discussed in section~4 below.

This model relies on the existence of
light scalars ($H'$ and the Higgs), and so must be
embedded in a framework that makes such scalars natural.
To embed this model into a supersymmetric model,
we add two additional ``Higgs'' chiral multiplets $H'_{u,d}$,
as well as $SU(2)_W$ triplets $\De_{u,d}$ with $Y = \mp 1$.
The relevant terms in the superpotential are
\beq[LeffSUSY3]\bal
\De W &\sim L H_u' \bar{X}
+ \De_u (H_u^2 + H'^2_u)
+ \De_d (H_d^2 + H'^2_d)
\\
&\qquad
+ \bar{X} X + H_u H_d + H'_u H'_d
+ \De_u \De_d
\eal\eeq
Integrating out $\De_{u,d}$ then generates \Eq{LeffnonSUSYUVcomplete}.
The mass terms may arise from the VEV of a singlet,
as discussed previously.
These interactions preserve both an $R$ parity under which the
$X$ fermion is even, and a $Z_2$ symmetry under which
$X$, $\bar{X}$, and $H'_{u,d}$ are odd and all other fields are
even.
It is therefore possible that this model contains a stable
LSP in addition to the dark matter particle.
In this case, the relic density of the LSP is constrained to be 
less than the observed dark matter density.
Alternatively, one of the $Z_2$ symmetries can be broken
by interactions such as
$\De W \sim X e^c (H_u H_d) (H'_u H_d)$,
allowing the LSP to decay.

\section{Direct Detection}
Asymmetric dark matter requires only very weak interactions with
the visible sector to explain the dark matter asymmetry,
and so there is no guarantee of an observable direct detection
cross section.
However, the symmetric component of the dark matter relic
density must be efficiently annihilated away, and if this
annihilation goes into standard model fields, the interactions
responsible for the annihilation can give rise to direct detection
similar to WIMP dark matter.

In previous section we
have presented two different minimal possibilities for the
interaction that
annihilates away the symmetric part of the dark matter, and each
of them leads to a different possible direct detection mechanism.
\begin{itemize}
\item
The annihilation may go through a singlet Higgs field whose
VEV gives the dark matter mass.
In this case the mixing between the singlet and the doublet
Higgs fields couples the dark matter to nucleons, giving
a potential direct detection signal several orders of magnitude
below current bounds.
\item
The annihilation may proceed via $t$-channel exchange of a
doublet scalar at the weak scale that is part of the UV
completion of the interaction that transfers the asymmetry.
In this case, there is a magnetic moment coupling of the dark
matter to nucleons that is closer to current bounds.
\end{itemize}
\noindent
We now discuss these possibilities in turn.

We first discuss singlet Higgs exchange.
The coupling \Eq{NMSSM} gives a coupling of the lightest
scalar Higgs
\beq
g_{\bar{X}X h^0} \sim \la_X \sin\th,
\eeq
where $\th$ is a Higgs-singlet mixing angle.
For a SM Higgs coupling to the nucleon, this gives a spin-independent
dark matter-nucleon cross section
\beq
\si_{\rm exp}(Xn \to Xn)
&= \frac{1}{A^2}
\frac{m_{Xn}^2}{m_{XN}^2} \si(XN \to XN) 
\nonumber\\
\label{spinindependentHiggsexsec}
&\simeq  5 \times 10^{-43}~\mbox{cm}^2 
\times g_{\bar{X}X h^0}^2
\left( \frac{m_{h^0}}{100\GeV} \right)^{-4}
\eeq
Here $\si_{\rm exp}$ is the experimentally quoted dark matter-nucleon
cross section and $m_{xy}$ is the reduced mass
(see {\em e.g.} \Ref{SilkReview}).
The best bound on this cross section for $m_X \simeq 15\GeV$
comes from XENON \cite{XENON},
$\si_{\rm exp} \lsim 9 \times 10^{-44}~\mbox{cm}^2$,
with a similar bound from CDMS \cite{CDMS}.
Since $\la_X = \sqrt{2} m_X / s \sim 0.1$ for $s \sim v$ near the
weak scale, 
the direct detection rate is about 2 orders of magnitude
below current sensitivity in this mass range.

In this model, the lightest pseudoscalar $a$ is light,
and its exchange gives rise to an effective coupling
of dark matter to quarks
\beq
\De\scr{L}_{\rm eff} = \frac{m_X}{s m_a^2} 
\bar{X} i \ga_5 X
\left[ 
\sum_u \frac{m_u \cot\be}{v} \bar{u} i\ga_5 u
+ \sum_d \frac{m_d \tan\be}{v} \bar{d} i\ga_5 d
\right].
\eeq
This gives a spin-dependent coupling to nucleons that is well
below current experimental bounds. 

We now turn to direct detection signals from the interaction
that transfers the asymmetry.
For the $L = \frac 12$ and $L = 1$ model the minimal UV
completion involves a scalar doublet, and if it is light
it can give sufficient annihilation.%
\footnote{The UV completion of the $B = \frac 12$ model
also involves electrically charged fields that may be 
light and lead to annihilation and direct detection.
We will not discuss this
possibility here.}
In this case, there is a one-loop magnetic moment and charge
radius coupling that can contribute to direct detection.
The magnetic moment and charge radius 
(defined in terms of the standard electromagnetic form
factors by $F_1(q^2) = q^2 r^2 / 6 + \cdots$,
$F_2(q^2) = \mu / 2 m_X + \cdots$)
were computed in \Ref{RabyWest}:
\beq
\mu &= \frac{y'^2}{32\pi^2} \frac{m_X^2}{m_{\phi^\pm}^2},
\\
r^2 &= -\frac{y'^2}{288 \pi^2} \frac{1}{m_{\phi^\pm}^2}
\left[ \ln \frac{m_{\phi^\pm}}{m_\tau} - \frac 34 \right],
\eeq
where $\phi^\pm$ is the charged scalar in the doublet
and we use their definitions of $\mu$ and $r^2$.
The charge-radius contribution to the cross section is 
IR divergent.
This is regulated using the ``energy transfer'' nuclear
cross section, which is given by \cite{RabyWest}
\beq[magmomcross]
\si(XN \to XN) = 
\frac{ \pi Z^2 \al^2}{m_X^2} \left[ \mu^2
+ \left( \frac{m_N}{m_N + m_X} \right)^2 
\left( 24 r^2 m_X^2 - \mu \right)^2 \right],
\eeq
where $m_N$ is the nucleus mass.
The cross section is numerically
dominated by the charge radius term.
Note that $\si \propto m_X^2$, so that $\si_n^{\rm exp}$
is independent of $m_X$ for $m_X \ll m_N$.
We obtain
\beq
\si_n^{\rm exp} \simeq 8 \times 10^{-44}~\mbox{cm}^2 
\left( \frac{Z/A}{0.4} \right)^2
\left( \frac{m_{\phi^\pm} / y'}{100\GeV} \right)^{-4}.
\eeq
This is below current bounds, but may give a
signal in upcoming experiments.

Finally, we mention that it is also possible to have
direct detection from interactions that are not directly
motivated by the physics of generating the dark matter
asymmetry.
For example, a $Z'$ mediator can also produce spin-independent
cross-sections of this size
\beq
\si_n^{\rm exp} \simeq 10^{-41} \mbox{ cm}^2
\left( \frac{g_{\bar{X}X Z'} g_{\bar{u} u Z'}}{10^{-1}} \right)^2
\left(\frac{1 \mbox{ TeV}}{M_{Z'}}\right)^4
\eeq  
This corresponds to a WIMP mass detected by DAMA of $\approx 7 \mbox{ GeV}$ \cite{Petriello,Savage,otherfits}, which results naturally from the models discussed above.

\section{Collider Signals}
\label{collide}

The new sector in ADM includes new states at the weak scale and below,
and thus have the potential to affect collider phenomenology.
The most significant possibilities are:
\begin{itemize}
\item New Higgs boson decays,
both invisible and into lighter decaying scalars.
\item Supersymmetry with NLSP decay resulting
in reduced missing energy, more leptons and/or new displaced vertices.
Kinematic shapes in cascade decays will also differ from standard
scenarios.
\item New charged states at the weak scale and and/or colored states at a TeV.
\end{itemize}
These signals can coexist in a single model.
We now discuss them in turn.

We begin with Higgs phenomenology.
If the NMSSM explains the dark matter annihilation,
then the Higgs has phenomenologically interesting couplings
to both the dark matter and the light pseudoscalar into which
the dark matter annihilates.
The couplings $\lambda_H$ and $\kappa$ must be ${\cal O}(1)$ to allow
for efficient enough annihilation.
The mixing angle between the lightest CP even Higgs and the scalar singlet
$s$ is 
\beq
\label{mixing}
\sin\theta_{s h^0} \simeq \frac{2\mu v}{s^2}\sin{2\beta} \sim \sin{2\beta}
\eeq
where $\mu$ is the effective mu-term, and the last approximation 
is due to annihilation requirements and chargino bounds.
Thus, the Higgs can decay to $X\bar{X}$ by mixing with $s$ with a decay width
\beq
\frac{\Ga(h\rightarrow X\bar{X})}{\Ga(h\rightarrow b\bar{b})}\simeq \frac{\lambda_X^2 \sin^2{2\beta} v^2}{3 m_b^2} .
\eeq
With $\lambda_X\sim 0.1$ to produce the correct $X$ mass,
and using the value of $m_b$ at the electroweak scale,
we find that this new decay is competitive at moderate $\tan\beta$,
implying a large invisible width for the Higgs. 

In addition, the Higgs can decay into pairs of light
pseudoscalars in this model.
This can dominate the Higgs width, especially since the couplings
$\la_H$ and $\ka$ are $\scr{O}(1)$.
The light pseudo-scalar mixes with the heavy CP-odd Higgs, $A^0$ and 
decays into $b\bar{b}$ or $\tau\bar{\tau}$,
thus producing a dominant decay of $h\rightarrow 4b$ or
$h\rightarrow 4\tau$ (for a review, see \cite{NShiggsRev}).
Of course a large partial width into these modes will suppress
the invisible decay mode discussed above.

We now turn to the decay of the would-be LSP in supersymmetric
theories.
This occurs in the $L = \frac 12$ and $B = \frac 12$ models
described above.

There are a large number of possible decay modes depending on
the model and the identity of the LSP.
These have certain common features that we point out before
discussing the individual cases.
As discussed above, in these models
the usual LSP is not the lightest particle
charged under a discrete $R$ symmetry.
Instead, there is a discrete $Z_4$ symmetry that allows
the would-be LSP to decay to \emph{pairs} of dark matter particles.
(We will refer to the would-be LSP as the NLSP.)
This means that LSP-mass reconstruction techniques \cite{Davis}
must be generalized to determine the mass of the dark matter.
Also, the decays are suppressed both by the scale $M$ in the
higher-dimension operators \Eq{LeffSUSY1} and \eq{LeffSUSY2},
and by the fact that the decays are often many-body decays.
This leads to the possibility of displaced vertices in the decays.
These operators have nontrivial flavor structure, and
considerations of approximate flavor symmetry suggest that they
are largest for heavy flavors;
if so, this leads to LSP decays involving heavy flavors,
which may be tagged.
The collider physics of these models is therefore extremely
rich and interesting.
In this paper, we will give only a sample of possible dominant
LSP decays, leaving detailed investigation for future work.

We begin with the $L = \frac 12$ model.
There are various possibilities for the identity of the NLSP.
If the NLSP is a neutralino, it has the decay
$\chi^0 \to \nu \bar{X}\bar{X}$ or $\bar{\nu} XX$
via a virtual sneutrino.
This is unfortunately completely invisible, and so gives
no modification of standard LSP phenomenology even for
macroscopic decay length.
It is important to note, however, that the reconstructed LSP
is not the dark matter particle in this case.
If the neutralino is sufficiently heavy,
there are also the 4-body visible decays
$\chi^0 \to h^0 \nu \bar{X}\bar{X}$,
$h^0 \bar{\nu} XX$,
$h^+ \ell^- \bar{X}\bar{X}$,
and $h^- \ell^+ XX$.
These decays proceed through a virtual left-handed slepton
like the dominant decay mode, so the branching ratio is
\beq
\mbox{BR}(\chi^0 \to h^0 \nu \bar{X} \bar{X}) 
\sim \frac{1}{8\pi^2}
\left( \frac{m_{\chi^0}}{v_u} \right)^2.
\eeq
This can easily be $\sim 1\%$ or larger,
and provide a window into LSP decay in this model.
The decay length for the visible decays is
\beq[neutLdecayLhalf]
c \tau(\chi^0 \to h^0 \nu \bar{X}\bar{X})
\sim \mbox{mm} 
\left( \frac{M}{10^6\GeV} \right)^2
\left( \frac{m_{\tilde{\nu}}}{200\GeV} \right)^4
\left( \frac{m_{\chi^0}}{100\GeV} \right)^{-7}.
\eeq
We see that the decay vertex is displaced for $M \gsim 10^4\GeV$.

If the $X$ scalars are lighter than the MSSM neutralino,
we have the decays
$\chi^0 \rightarrow \nu \tilde{X} \tilde{X}$,
$\bar\nu \bar{\!\tilde X}\gap\gap \bar{\!\tilde X}$.
These proceed directly through the interaction \Eq{LeffSUSY1}
without a virtual intermediate state, and are therefore
enhanced compared to $\chi^0 \to \nu \bar{X}\bar{X}$.
The decay length is
\beq[neutXdecayLhalf]
c \tau(\chi^0 \to \bar{\nu} \tilde{X}\tilde{X})
\sim \mbox{cm}
\left( \frac{M}{10^8\GeV} \right)^2
\left( \frac{m_{\chi^0}}{100\GeV} \right)^{-3},
\eeq
The $X$ scalars subsequently decay via
$\tilde{X} \rightarrow \bar{X} \nu$,
which is a completely invisible mode, or the subleading visible decay mode
$\tilde{X} \rightarrow X \nu h^0$,
$\tilde{X} \rightarrow X \ell^{\pm} h^{\mp}$,
assuming it is kinematically available.
The branching ratio for the visible mode is suppressed only
by 3-body phase space, and so can have a branching ratio
of up to $\sim 1\%$ if the 3-body channel is fully open.
The decay length for the visible mode is approximately the same as
\Eq{neutXdecayLhalf}.
The general pattern for $\chi^0$ decays
is therefore qualitatively the same whether the
scalar decay channel is open or not:
the dominant decay is invisible, possibly with rare decays to 
Higgs plus missing energy, or charged Higgs plus charged lepton and missing energy.

We now consider the case where the NLSP is a slepton or squark.
In this case, the dominant decay is through a virtual gaugino:
$\tilde{q} \to q \tilde{\chi}^*$
or $\tilde{\ell} \to \ell \tilde{\chi}^*$.  
The virtual gaugino then decays through the same modes as given for the real gaugino above.

For example, suppose the NLSP is a right-handed stau.
(We expect the heaviest flavor slepton or squark to be the lightest, 
since the Yukawa couplings drive down the scalar masses
in the renormalization group equation.)
This can occur, for example, in models of gauge-mediated SUSY breaking.
If the $X$ scalars are heavy, the dominant decay mode is
$\tilde{\tau}_R \to \tau \nu \bar{X}\bar{X}$,
$\tau \bar\nu XX$
via a virtual neutralino.
The decay length is
\beq[tauRdecayLhalf]
\!\!\!\!\!\!\!\!
c \tau(\tilde{\tau}_R \to \tau \nu \bar{X}\bar{X})
\sim \mbox{mm} 
\left( \frac{M}{10^6\GeV} \right)^2
\left( \frac{m}{200\GeV} \right)^6
\left( \frac{m_{\tilde{\tau}}}{100\GeV} \right)^{-7},
\eeq
where we have assumed a common mass scale
$m \sim m_{\tilde{\nu}} \sim m_{\chi^0}$.
The chargino couples to the right-handed stau only through
the tau Yukawa, but this may be important at large $\tan\be$.
Virtual chargino exchange gives the decays
\beq
\mbox{BR}(\tilde{\tau}_R^- \to \nu_\tau \ell^- \bar{X}\bar{X})
\sim 10^{-4} \tan^2\be
\left( \frac{m_{\chi^0}}{m_{\chi^\pm}} \right)^2,
\eeq
which may involve light leptons, depending on the
flavor structure of the interaction \Eq{LeffSUSY1}.

As for the neutralino NSLP, decays to $X$ scalars are dominant
if they are kinematically allowed.
The decays are  
$\tilde{\tau}_R^- \to \ell \nu \tilde{X} \tilde{X}$ with decay length
\beq[tauRdecayScalarLhalf]
\!\!\!\!\!\!\!\!
c \tau(\tilde{\tau}_R \to \ell \nu \tilde{X}\tilde{X})
 \sim \mbox{cm} 
\left( \frac{M}{10^7\GeV} \right)^2
\left( \frac{m}{200\GeV} \right)^2
\left( \frac{m_{\tilde{\tau}}}{100\GeV} \right)^{-5},
\eeq
This is followed by $\tilde{X} \to \bar{X} \nu$
with decay length given by \Eq{neutXdecayLhalf}.

The right-handed stau will mix with the left-handed stau
via $A$ terms and the $\mu$ term times the tau Yukawa coupling.
It is also possible that the left-handed stau is the LSP.
We therefore consider the decays of a left-handed stau.  If kinematically available, the most important decay is directly through the operator \Eq{LeffSUSY1}, $\tilde{\tau}_L \to h^- X X$.  The decay length is given by \Eq{neutXdecayLhalf}.  If this channel is not open,
the leading decay is $\tilde{\tau}_L \to W XX$ via a
virtual snuetrino.
The decay rate is parametrically the same as \Eq{tauRdecayLhalf}.
For sufficiently large $m_{\tilde{\tau}}$, we also
have the decay $\tilde{\tau}_L^- \to XX \bar{t} b$
and $\tilde{\tau}_L^+ \to \bar{X}\bar{X} \bar{b} t$.
The ratio of decay rates is
\beq
\frac{\Ga(\tilde{\tau}_L^- \to XX \bar{t} b)}
{\Ga(\tilde{\tau}_L^- \to W^-XX)}
\sim 0.1 \left( \frac{m_{\tilde\nu}}{m_{H^\pm}} \right)^4
\left( \frac{m_{\tilde\tau}}{300\GeV} \right)^2.
\eeq
This may be the dominant decay, depending on the
superpartner masses.

Returning to the case of right-handed stau LSP, the mixing
with $\tilde{\tau}_L$ allows the decay to the final states
discussed for $\tilde{\tau}_L$ decay above:
\beq
\De\Ga(\tilde{\tau}_R \to \cdots)
\sim \left( \frac{m_\tau \mu \tan\be}{m^2} \right)^2 
\Ga(\tilde{\tau}_L \to \cdots)
\eeq
Therefore, $\tilde{\tau}_R \to h^- X X$ may be an important decay mode in some models, particularly at large $\tan\be$.

If a squark (\eg\ the stop) is the LSP, then the 
dominant decay modes are
$\tilde{q} \to q \nu \bar{X}\bar{X}$,
$q \bar{\nu} XX$ 
via a virtual neutralino,
or $\tilde{q} \to q \ell \bar{X} \bar{X}$,
$q \bar{\ell} XX$ 
via a virtual chargino.
The latter decay will be suppressed if the LSP is 
dominantly a right-handed squark.
The decay length is as in \Eq{tauRdecayLhalf}
with the obvious replacements.

We now turn to the $B = \frac 12$ model.
In general, the decay lengths in the $B = \frac 12$ model are
longer than the $L = \frac 12$ model because the operator
\Eq{LeffSUSY2} is suppressed by a higher power of $M$.
In addition, some decays are loop suppressed.
We begin with the case of neutralino NSLP.
If the $X$ scalars are heavy, the neutralino will decay via
$\chi^0 \to udd\, \bar{X}\gap\gap \bar{X}$ or
$\bar{u} \bar{d} \bar{d} {X}{X}$.
The diagram involves a virtual quark connecting to $X$ fermions,
with the remaining squarks converting to quarks via a loop. 
The decay length for this process is
\beq
c\tau(\chi^0 \to X X q q q) \sim 100~\mbox{m}
\left( \frac{M}{\mbox{TeV}} \right)^4
\left( \frac{m}{500\GeV} \right)^{6}
\left( \frac{m_{\chi^0}}{100\GeV} \right)^{-11}.
\eeq
This estimate is highly uncertain due to the estimate of 4-body
phase space and the loop factor.
For such large displaced vertices
the collider physics of the SUSY LSP is unchanged,
but the LSP reconstructed at colliders is not the dark matter.
As above, decay to the $X$ scalars is the
preferred mode when it is kinematically allowed.
The decay modes are
$\chi^0 \to udd\, \bar{\!\tilde X}\gap\gap \bar{\!\tilde X}$
or $\bar{u} \bar{d} \bar{d} \tilde{X}\tilde{X}$
followed by by
$\,\bar{\!\tilde X}\gap\gap \to X \bar{u} \bar{d} \bar{d}$
or $\tilde{X} \to \bar{X} udd$.  
The decay length to $X$ scalars is
\beq
c\tau(\chi^0 \to qqq\tilde{X}\tilde{X}) \sim 0.3~\mbox{mm}
\left( \frac{M}{\mbox{TeV}} \right)^4
\left( \frac{m}{500\GeV} \right)^4
\left( \frac{m_{\chi^0}}{100\GeV} \right)^{-9}.
\eeq
This is followed by the $\tilde{X}$ decay with
\beq[BhalfXdecay]
c\tau(\tilde{X} \to X q q q) \sim 3~\mbox{mm}
\left( \frac{M}{\mbox{TeV}} \right)^4
\left( \frac{m}{500\GeV} \right)^{2}
\left( \frac{m_{\tilde{X}}}{100\GeV} \right)^{-7}.
\eeq
We again emphasize that this involves crude estimates of the 
many-body phase space and loop factors.
There are additional complications from the flavor structure of the interaction.
If the dominant terms in \Eq{LeffSUSY2} involve right-handed tops,
there will be additional suppression if $\chi^0$ decays to tops
are kinematically forbidden.

We now move onto the case of a squark LSP.  If the $X$ scalars are light enough, the decay will go directly through the operator \Eq{LeffSUSY2} with decay length
\beq
c\tau(\tilde{q} \to \tilde{X} \tilde{X}  q q) \sim 10^{-8}~\mbox{m}
\left( \frac{M}{\mbox{TeV}} \right)^4
\left( \frac{m_{\tilde{q}}}{100\GeV} \right)^{-5}.
\eeq
The $X$ scalars subsequently decay via \Eq{BhalfXdecay}.
If the squark decay to $X$ scalars is not kinematically available,
the decay must go through a squark loop with decay length
\beq
c\tau(\tilde{q} \to X X q q) \sim 0.3~\mbox{cm}
\left( \frac{M}{\mbox{TeV}} \right)^4
\left( \frac{m}{500\GeV} \right)^{2}
\left( \frac{m_{\tilde{q}}}{100\GeV} \right)^{-7},
\eeq
where $m$ is an assumed common mass in the loop.

Finally we consider the case of a slepton LSP.
The decay must go via a virtual gaugino $\tilde{\ell} \rightarrow \ell \chi^*$.  The virtual gaugino has the same decay modes as the real gauginos discussed above.  Thus through a neutralino, we may have  $\tilde{\ell} \to \ell q q q XX$ with decay length
\beq
c\tau(\tilde{\ell} \to \ell q q q XX) \sim 10^5~\mbox{m}
\left( \frac{M}{\mbox{TeV}} \right)^4
\left( \frac{m}{500\GeV} \right)^{8}
\left( \frac{m_{\tilde{\ell}}}{100\GeV} \right)^{-13}.
\eeq
If the $X$ scalars are light enough, the slepton can decay 
via $\tilde{\ell} \to \ell q q q \tilde{X}\tilde{X}$ with decay length
\beq
c\tau(\tilde{\ell} \to \ell q q q \tilde{X}\tilde{X}) \sim 1~\mbox{m}
\left( \frac{M}{\mbox{TeV}} \right)^4
\left( \frac{m}{500\GeV} \right)^{6}
\left( \frac{m_{\tilde{\ell}}}{100\GeV} \right)^{-11}.
\eeq
Again, these are highly uncertain estimates.

In several of the scenarios discussed above,
the $X$ fermions and/or scalars can have lifetimes on
cosmological time scales.
This may be due either to larger values of $M$, different
superpartner masses,
or simply because of large suppression factors missed in the crude
approximations made above. 
Such decays can have important effects on
nucleosynthesis, matter-radiation equality,
or the dark-matter content of the universe, among other issues.
The study of these issues
is beyond the scope of this work.

Finally, a few comments on the $L=1$ model.
In this case the operator \Eq{LeffSUSY3} does not break
$R$-parity so that both the LSP and the dark matter particle may be stable.
In the non-SUSY UV completion of this model,
there is some novel phenomenology associated with the $H'$ state.
As already mentioned in Sec.~2.3, the charged Higgs can be pair produced.
They then decay via $H'^{\pm} \to \tau^\pm + \met$,
the missing energy being carried away by dark matter.

\section{Conclusions}
We have presented a simple class of models in which the dark matter relic
abundance is determined by the baryon asymmetry.
This naturally explains the fact that the observed baryon and dark matter
abundances are close, $\Om_{\rm DM} \simeq 5 \Om_{\rm B}$.
Conceptually these models are very simple: higher dimension operators
distribute the primordial $B-L$ asymmetry between the dark and visible sectors.
When these higher dimension operators fall out of equilibrium, the asymmetry
is separately frozen into the two sectors.  
We presented several simple examples as existence proofs.
In any specific model, the dark matter mass is precisely determined,
and we find masses in the range $5$ to $15\GeV$ in the models presented here.

The symmetric component of the dark matter abundance can be annihilated
away either by some of the same interactions that transfer the asymmetry,
or by the interactions that generate the dark matter mass.
Interestingly, either of these mechanisms also gives a possible direct
detection cross section large enough to be observed in upcoming experiments.
This gives a strong motivation for additional dark matter experiments and analyses
sensitive to dark matter in the low mass range.
An interesting feature in the second class of models is that the
dominant direct detection mechanism is through the electric charge
radius of the dark matter.

This dark matter mechanism described here
can be naturally combined with supersymmetry,
giving a connection to a plausible model of electroweak physics
and the hierarchy problem.
In some models, the would-be LSP (and dark matter candidate) of
supersymmetry naturally decays into \emph{pairs} of dark matter
particles.
This gives rise to interesting modifications of standard
supersymmetric collider phenomenology, including the possibility
of highly displaced vertices.

This class of models gives a compelling alternative
to the usual WIMP paradigm for dark matter that is worthy of 
further investigation.

\section*{Acknowledgements}
We thank G.\ Kribs, A.\ Kusenko and J.\ Terning for discussions.
We thank the Kavli Institute for Theoretical Physics
and the Aspen Center for Physics,
where part of this work was performed.
This work was supported by the National Science Foundation under grant NSF-PHY-0401513 (DEK) and
by the US Department of Energy, including grant DE-FG02-95ER40896
and by NASA grant NAG5-18042 (KMZ).

\newpage

\end{document}